\documentclass[twocolumn,aps,showpacs,preprintnumbers,amsmath,amssymb]{revtex4}

\usepackage{graphicx}
\usepackage{dcolumn}
\usepackage{bm}
\usepackage{graphics}
\usepackage{longtable}

\begin{document}


\title{Atom counting in ultra-cold gases using photoionisation and ion detection}

\author{T. Campey}
\email{t.campey@uq.edu.au}
\author{C. J. Vale}
\author{M. J. Davis}
\affiliation{School of Physical Sciences, The University of
Queensland, Brisbane, QLD 4072, Australia}
\author{S. Kraft}
\author{C. Zimmermann}
\author{J. Fort\'{a}gh}
\affiliation{Physikalisches Institut der Universit\"{a}t
T\"{u}bingen, 72076 T\"{u}bingen, Germany}
\author{N. R. Heckenberg}
\author{H. Rubinsztein-Dunlop}
\affiliation{School of Physical Sciences, The University of
Queensland, Brisbane, QLD 4072, Australia}


\pacs{03.75.-b, 03.75.Be, 32.80.Fb, 39.90.+d}

\begin{abstract}
We analyse photoionisation and ion detection as a means of
accurately counting ultra-cold atoms.  We show that it is possible
to count clouds containing many thousands of atoms with accuracies
better than $N^{-1/2}$ with current technology. This allows the
direct probing of sub-Poissonian number statistics of atomic
samples. The scheme can also be used for efficient single atom
detection with high spatio-temporal resolution.  All aspects of a
realistic detection scheme are considered, and we discuss
experimental situations in which such a scheme could be
implemented.
\end{abstract}

\maketitle

\section{Introduction}

The term ``atom optics'' arises from the analogy between
experiments exhibiting the wave-like properties of light with
those demonstrating the wave-like properties of matter
\cite{adams}. Due to the vanishingly small de Broglie wavelength
of room temperature atoms it is necessary to cool gases to
microKelvin temperatures before these properties are observable.
Before 1995, experiments in atom optics used clouds of
laser-cooled thermal atoms that can be considered similar to
thermal light sources in optics, with coherence lengths of less
than a micron.

The realisation of Bose-Einstein condensation in 1995
\cite{Anderson95,Bradley95,Davis95} provided the first intense
sources of coherent matter waves, analogous to the invention of
the optical laser.  Many of the early experiments with BECs
concentrated on their wave-like properties, which in general can
be well described by the Gross-Pitaevskii equation for the
classical field $\psi(x) \approx \langle \hat{\Psi}(x) \rangle$.
Some notable experiments include the interference of independently
prepared condensates \cite{andrews97} and the demonstration of
atom lasers \cite{mewes97,bloch99,hagley99,lecoq01}. These
experiments rely on the first-order coherence of the condensate,
and in some sense can be considered to be coherent ``classical
atom optics''.

Recently there have been an increasing number of experiments
focusing on the generation and measurement of non-trival higher
order coherences in matter waves that are beyond the scope of
Gross-Pitaevskii mean field theory.  These experiments could be
said to be in the area of ``quantum atom optics''
\cite{moelmer03}.  The development of this new area bears a close
resemblance to the development of quantum optics following the
invention of the optical laser.  Just as quantum optics relies on
single photon detectors, quantum atom optics will utilise
detectors with single atom resolution, such as the technique we
describe in this paper.

One of the earliest experiments probing higher-order coherence in
BECs was in the measurement of three-body loss from a BEC, from
which the correlation function $g^{(3)}(\mathbf{x})$ could be
inferred \cite{burt97}.  There have been several measurements made
of the local second-order correlation function $g^{(2)}(\mathbf{x})$
for a number of systems, including a 1D strongly interacting Bose
gas \cite{tolra04, kinoshita05}, a quasi-condensate \cite{esteve06},
and a thermal Bose gas \cite{schellekens05}. The value of
$g^{(2)}(\mathbf{x})$ for an atom laser has been measured in ref.
\cite{ottl05}.

There has also been strong interest in relative number squeezing.
This is where a state is generated in which the number difference
between two or more atomic samples is sub-Poissonian. Javanainen
and Ivanov \cite{javanainen99} showed that splitting condensates
by adiabatically turning up a tunnel barrier can produce number
squeezed atomic states. Experimentally, Orzel \emph{et al.}
\cite{orzel} loaded a BEC into a one-dimensional optical lattice
and observed the degradation of interference fringes. This was
interpreted to be caused by increased phase-fluctuations due to
reduced number fluctuations at the lattice sites.  A similar
observation was made in the first demonstration of a 3D Mott
insulator state \cite{greiner02} and further elucidated by other
related experiments \cite{hadzibabic04}.

Correlations and relative number squeezing have been predicted in
collisions and four-wave mixing in condensates
\cite{Pu00,Duan00,Sorensen01,roberts02}, as well as in
downconversion of a molecular BEC  in both the spontaneous
\cite{kherunts02} and stimulated  \cite{kherunts05} regimes. While
experiments have observed four-wave mixing processes \cite{Deng1999,campbell06}
and matter-wave amplification \cite{inouye99}, there has been no
direct measurement of these correlations.

Experimentally, other high-order correlations have been directly
detected by suitable analysis of absorption images from an ensemble
of experiments \cite{greiner05,folling05}. Other schemes such as the
``quantum tweezer'' \cite{diener02} can deterministically extract
sub-Poissonian atomic samples from a BEC. Optical dipole traps along
with fluorescence imaging have been used to demonstrate
sub-Poissonian statistics in small samples of atoms \cite{chuu05}.

In order to directly probe atom number statistics of cold quantum
gases, one would ideally like an atom counter with accuracy at the
single atom level.  Existing single atom detection schemes can be
divided into two categories: optical and contact. One of the
simplest optical schemes is based on resonance fluorescence
detection \cite{chuu05,frese00,schlosser01}. Assuming negligible
background, the shot noise limited atom number uncertainty in a
fluorescence measurement scales as $\sqrt{N/\alpha\gamma\tau}$,
where $N$ is the number of atoms, $\alpha$ is the collection
efficiency of the detector, $\gamma$ is the mean scattering rate and
$\tau$ is the integration time \cite{teper06}. With a long enough
integration time it is possible to measure atom numbers with
sub-Poissonian precision. However, limitations such as atom heating
restrict the capability of sub-Poissonian fluorescence detection to
atom numbers of order $<$ 100 \cite{chuu05}. Absorption measurements
offers more favourable scaling to higher atom numbers \cite{teper06}
but achieving good absolute accuracy remains technically difficult.
High finesse optical cavities offer excellent sensitivity for single
atom detection \cite{hood00,pinske00,mabuchi02} and can be used for
slow counting of a large number of atoms. Cavities offer an
improvement in signal to noise of $\sqrt{\mathcal{F}}$, where
$\mathcal{F}$ is the cavity finesse, compared to single pass
measurements for a fixed amount of heating \cite{horak03,lye03}.
Moderate cavities have also been shown to aid fluorescence and
absorption measurements \cite{teper06}.  It is difficult however to
apply cavity techniques to counting large atom numbers and long
integration times are required.

The second category of single atom detectors is based on contact
methods. These generally involve a charged particle or metastable
atom with high internal energy impacting on a charged particle
detector. This initiates a cascade of electron emissions, which are
amplified to produce a macroscopic current pulse. This method is
particularly useful for detecting metastable He atoms
\cite{baldwin05} where the arrival statistics of atoms released from
a BEC allowed for matter wave Hanbury-Brown Twiss correlations to be
observed \cite{schellekens05}. While this detection method is
simple, metastable atoms remain difficult to condense and most cold
atom experiments use alkali atoms.

In this paper we describe an atom detection scheme based on
photoionisation and ion detection. This scheme overcomes some of the
difficulties associated with optical detection, can be applied to
ground state alkali atoms and is capable of extending sub-Poissonian
counting to much larger numbers of atoms. Photoionisation
\cite{ruschhaupt04} and ionisation \cite{ristrophe05} have
previously been considered as a means of efficient atom detection.
However, these proposals have not included many factors present in
realistic detectors. Firstly, we note that the best absolute
detection efficiencies of room temperature charged particle
detectors are of order 0.8--0.9. (Superconducting tunnel junctions
can in principle provide unity detection efficiency \cite{frank96}
but are impractical for many experimental setups.) This has serious
consequences for any experiment which endeavours to count a large
number of atoms, but as we show in this paper, provided the
detection efficiency is greater than 0.5, it is still possible to
count atoms with accuracies better than $1/\sqrt{N}$. The achievable
accuracy depends on several factors including the detector
calibration, background count rates, detector pulse resolution
times, ionisation rates and the uncertainties in these parameters,
all of which are included in our analysis.

This paper is organised as follows: in Section II we describe the
experimental arrangement necessary for atom counting and how it can
be applied to a cloud of trapped $^{87}$Rb atoms. Section III
addresses the statistics associated with the detection scheme along
with the requirements on the detector efficiency. In Section IV we
discuss detector calibration and describe numerical simulations of
the calibration including the effects mentioned above. Section V
looks at the application of the scheme in the presence of losses and
we follow with a discussion of potential applications. Appendix
\ref{app:symbols} contains a table of symbols and Appendices
\ref{app:Nf} and \ref{app:pr} detail the derivations of certain
expressions used in the paper.

\section{Scheme}

The method we propose consists of slowly photoionising a sample of
atoms and accelerating the ions to a charged particle detector.
Figure 1(a) shows the elements of the scheme. A cloud of cold atoms
is held in an optical dipole trap where it is illuminated by the
ionisation lasers. Our discussion focusses specifically on the case
of $^{87}$Rb atoms but the scheme could be modified to detect other
alkali atoms by the correct choice of lasers. Figure 1(b) shows the
relevant energy levels and transitions for $^{87}$Rb.

\begin{figure}
\includegraphics[width=3.2in]{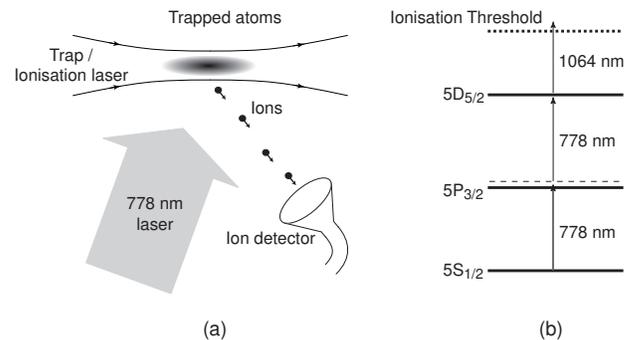} \caption{(a) In the proposed atom counting scheme atoms are ionised
from an optical dipole trap via a three-photon process. The
resulting ions are accelerated into an ion detector. (b) The
ionisation scheme showing the relevant energy levels of $^{87}$Rb
(not to scale). The scheme consists of 2 $\times$ 778 nm photon
excitation to the 5D$_{5/2}$ state, followed by ionisation with a
Nd:YAG laser, which also forms the dipole trap.} \label{fig:scheme}
\end{figure}

The ionisation scheme consists of two stages. The first is
two-photon excitation with a low intensity 778 nm laser from the
5S$_{1/2}$ state to the 5D$_{5/2}$ state. The second is ionisation
from this state with a high intensity infra-red laser such as a
Nd:YAG laser. The 5D$_{5/2}$ state has a photoionisation
cross-section approximately three orders of magnitude higher than
the ground state \cite{duncan01} so ionisation can readily be
achieved with a cw laser of wavelength shorter than \mbox{$\sim$
1200 nm} rather than with pulsed lasers \cite{ciampini02}. Ion
detection is achieved after appropriate ion collection optics using
a standard ion detector such as a channel electron multiplier (CEM)
or discrete dynode detector.

The atom counting scheme works best if the pulses produced by ions
striking the ion detector are well resolved in time.  This means the
ionisation rate $\dot{N}_I$ should be much lower than the inverse of
the pulse resolution time of the detection system $1/\tau_r$. In
Section V we will see that this means that for typical detector
efficiencies and trap lifetimes, and atom numbers of the order of
$10^3$ or $10^4$, the ionisation should take place on a time scale
of order several ms. In order to confine the atoms during this time,
the Nd:YAG laser is focused to create a dipole trap.  This localises
the atoms in the highest intensity region of the beam. The intensity
must be high enough to produce a trap sufficiently deep that no
atoms are lost due to heating as the ionisation takes place. The
major source of heating is the spontaneous decay of atoms excited to
the 5D$_{5/2}$ state. A high Nd:YAG laser intensity thus helps
eliminate trap loss not only by creating a deep trap, but also by
ensuring that an atom excited to the 5D$_{5/2}$ state has a
substantially higher probability of being ionised than of decaying
back to the ground state. Atoms that decay from the 5D$_{5/2}$ state
can end up in several ground state sublevels.  A benefit of the
dipole trap is that it traps all of these sublevels, in contrast to
magnetic fields, which can only trap a restricted class of
sublevels.

The rate of ionisation can be controlled with the 778 nm laser
intensity. In Section V we show that for a given atom number and
trap loss rate there is an optimal ionisation rate that minimises
the uncertainty in the inferred atom number. In order to hold the
ionisation rate approximately constant at the ideal value, the 778
nm laser intensity can be exponentially ramped to compensate for the
reduction in trapped atom number as ions are produced. However we
also show that the scheme works well with constant laser
intensities.

This state selective scheme also represents an efficient single atom
detector \cite{ionoptics}.  The ionisation lasers can be tightly
focussed to detect atoms at a well defined spatial location.  The
two photon excitation rate is proportional to the squared intensity
of the 778 nm laser so that scattered light is less likely to heat
atoms not in the detection region.  Additionally, for single atom
detection the spatial intensity profiles of the two lasers can be
chosen so that the repulsive optical dipole potential of the
blue-detuned 778 nm laser can be compensated for by the attractive
potential of the high power red-detuned ionising laser.  This leads
to minimal perturbation of undetected atoms which may for instance
be magnetically detuned or trapped in a different internal state.
These factors, and the relatively quick ionisation times, make this
a flexible scheme well suited to single atom detection with
excellent spatio-temporal resolution.

\section{Counting with imperfect detectors}

We now consider how accurately an atom number can be determined
using a scheme in which ions are counted with a non-unity efficiency
detector. The overall ion detection efficiency is
\begin{equation} \label{eq:eta} \eta_i = \eta_{det}
p_{r},
\end{equation} where $\eta_{det}$ is the efficiency of the ion detection system and is the product of the
ion detector efficiency $\eta'_{det}$ and the collection efficiency
$\eta_{coll}$. The quantity $p_{r}$ is the probability that an ion
is temporally resolvable from the other ions detected, and is
ionisation rate dependent.

If $N_I$ ions are produced, the number of ions counted, $N_i$, is
described by a binomial distribution.  Provided $\eta_i(1-\eta_i)
N_I$ is greater than $\sim 5$, this is well approximated by a
normal distribution with mean $\eta_i N_I$ and standard deviation
$\sqrt{\eta_i(1-\eta_i)N_I}$ \cite{knoll79}. In a given
experiment, the number of atoms ionised can be inferred to be
\begin{equation} \label{eq:NIinfbar} N_{I,inf}=\frac{N_{i}}{\eta_i},
\end{equation} with an uncertainty \begin{equation} \label{eq:NIinfsig}
\sigma_{N_{I,inf}}=\sqrt{\frac{N_i(1-\eta_i)}{\eta_i^2}+\frac{N_i^2
\sigma_{\eta_i}^2}{\eta_i^4}}, \end{equation} where
$\sigma_{\eta_i}$ is the uncertainty in $\eta_i$.

In order to compare this uncertainty to Poissonian fluctuations, we
define the normalised count uncertainty:
\begin{equation} \label{eq:F} \kappa \equiv
\frac{\sigma^2_{N_{I,inf}}}{N_{I,inf}}. \end{equation}

The quantity $\kappa$ is related to the measurement Fano factor
discussed in Section \ref{sect:discussion} but includes an
additional systematic contribution due to the uncertainty in the
detection efficiency (the second term under the square root in Eq.
\ref{eq:NIinfsig}). $\kappa = 1$ implies the count uncertainty is
equal to the uncertainty inherent in Poissonian statistics. In order
to count ions with error less than $\sqrt{N_I}$, $\kappa$ must be
less than one.

\begin{figure}
\includegraphics[width=3.5in]{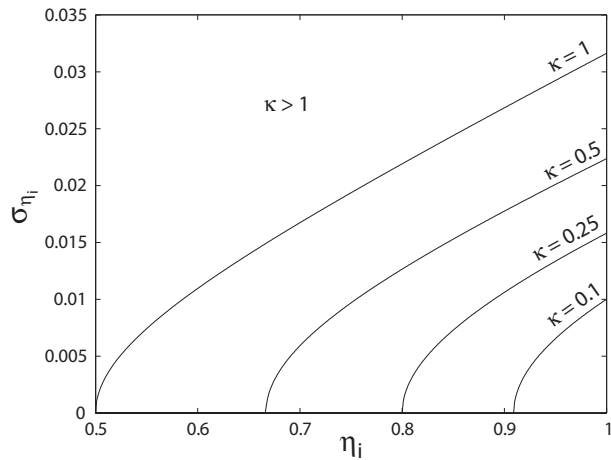} \caption{Contours of the normalised count uncertainty $\kappa$ for an atom counting system with ion detection
efficiency $\eta_i \pm \sigma_{\eta_i}$, for ion number $N_I=10^3$.}
\label{fig:pkplot}
\end{figure}

In Fig. \ref{fig:pkplot}, contours of the normalised count
uncertainty $\kappa$ are plotted as a function of $\eta_i$ and
$\sigma_{\eta_i}$ for $N_I=10^3$. From this figure it can be seen
that for a detection system in which the detection efficiency is
known exactly, i.e. $\sigma_{\eta_i}=0$, the overall ion detection
efficiency $\eta_i$ need only be greater that 50\% to achieve
$\kappa$ less than 1. As the uncertainty in the detection efficiency
increases, the required detection efficiency to achieve $\kappa\leq
1$ also increases. Thus, to achieve a low value of $\kappa$, we
require a high efficiency detector that is well calibrated. In the
next section we examine how this can be done.

\section{Detector calibration}

To infer the number of atoms ionised and its uncertainty from an ion
count, we need to know the overall ion detection efficiency $\eta_i$
and its uncertainty $\sigma_{\eta_i}$.  From Eq. (\ref{eq:eta}) this
means we need to know the ion detection system efficiency
$\eta_{det}$ and the ion temporal resolution probability $p_r$, each
with their respective uncertainties. We first consider how to
measure $\eta_{det}$ to high accuracy, or in other words, how to
calibrate the ion detection system.

Accurate calibration can be achieved by means of a scheme in which
an additional detector is used to count the electrons produced by
ionisation \cite{ncrp}. If $N_I$ atoms are ionised in a calibration,
and the overall ion and electron detection efficiencies are $\eta_i$
and $\eta_e$ respectively (where analogous to Eq. (\ref{eq:eta}),
 $\eta_e=\eta_{det,e}p_{r,e}$), then \begin{eqnarray} \label{eq:calib1} N_i&=& \eta_i N_I,\\
N_e&=&\eta_e N_I,\\ N_{c}&=&\eta_i \eta_e N_I, \end{eqnarray} where
$N_i$ is the number of ions counted, $N_e$ is the number of
electrons counted and $N_{c}$ is the number of ion-electron
coincidences counted. Therefore \begin{equation} \label{eq:calib4}
\eta_i=\frac{N_{c}}{N_e}, \end{equation} which shows that only the
coincidence and electron counts are required to find the ion
detection efficiency; we don't need to know the efficiency of the
electron detector.

In order to count the number of coincidences, it is first necessary
to specify how coincidences are identified.  In general there will
be a distribution of times taken for ions to move from the point of
ionisation to striking the detector and producing a pulse. The width
of this distribution will depend on the range of initial velocities
of the ions and the variation in electric field at different
locations within the atom cloud.  The same is true for the
electrons. There will therefore be a distribution of intervals
between ions and their corresponding electrons.  With modern pulse
counting electronics it is straightforward to produce lists of the
arrival times of electrons and ions with precisions on the order of
100 ps \cite{ortec} and these can be used to generate a list of
intervals between electrons and the following ions.  If a histogram
of these intervals is plotted, true coincidences will appear as a
peak against a small background of false coincidences. This peak can
then be used to define a window for true coincidences.

The coincidence window may also contain false coincidences, which
occur when an unpartnered ion (i.e. one for which the corresponding
electron was not detected) arrives within the window of an
unpartnered electron. Also, due to the finite size of the window,
some true coincidences may not be counted. The mean net number of
false coincidences $\bar{N}_f$ and its uncertainty $\sigma_f$ are
derived in Appendix \ref{app:Nf}.

In order to take false coincidences and background ions and
electrons into account in the ion detection efficiency, we define
corrected ion, electron and coincidence counts as $N_i'\equiv
N_i-\bar{N}_{ib}$, $N_e'\equiv N_e-\bar{N}_{eb}$ and $N_c'\equiv
N_c-\bar{N}_f$, respectively. $\bar{N}_{ib}$ and $\bar{N}_{eb}$ are
respectively the mean numbers of background ions and electrons
counted in the calibration period and can be measured by counting
ions and electrons in the absence of the calibration atoms.  The
overall ion detection efficiency is then
\begin{equation} \label{eq:etai} \eta_{i}=\frac{N_c'}{N_e'}. \end{equation}
The ion detection efficiency given by Eq. (\ref{eq:etai}) is the
value that applies in a calibration and is valid for the ionisation
rate used in the calibration. To find the system ion detection
efficiency (which is ionisation rate independent) we need to correct
$\eta_i$ for the value of the ion temporal resolution probability
$p_r$ that applies in the calibration. From Eqs (\ref{eq:eta}) and
(\ref{eq:etai}), the system ion detection efficiency is
\begin{equation} \label{eq:etadet} \eta_{det}=\frac{N'_c}{N'_e p_r},
\end{equation} and from Appendix \ref{app:pr} \begin{equation} \label{eq:pr}
p_r=1-\tau_r \dot{N}_i - \frac{1}{2}(\tau_r \dot{N}_i)^2,
\end{equation} where $\tau_r$ is the minimum time interval for which
two ions can be resolved from each other, and $\dot{N}_i$ is the
average rate of ion detection in the calibration.

We also need to determine the uncertainty in $\eta_{det}$. This
uncertainty arises mainly from the fact that for a given number of
atoms ionised, the number of ion, electron and coincidence counts
are described by binomial distributions.  An additional contribution
comes from the uncertainties in the number of background electrons
and false coincidences counted during the calibration.  Taking all
of these effects into account, and for values of $p_r$ close to
unity, the uncertainty in the ion detection system efficiency is
\begin{equation} \label{eq:sigetai} \sigma_{\eta_{det}}=
\frac{N_c'}{N_e'} \sqrt{\left(\frac{1}{N_c'}-\frac{1}{N_e'}\right)+
\frac{\sigma_{eb}^2}{N_e'^2}+\frac{\sigma_f^2}{N_c'^2}},
\end{equation} where $\sigma_{eb}$ is the uncertainty
in the number of background electrons counted and is equal to
$\sqrt{\bar{N}_{eb}}$, assuming the detection of background
electrons is a Poisson process.

To minimise the uncertainty in the ion detection efficiency, it is
desirable to minimise background counts. Background electrons and
ions arise from three main potential sources. The vacuum gauges and
ion getter pumps commonly used in ultra-cold atom experiments work
by ionising atoms and hence some of the ionisation products may find
their way to the detector. While the vacuum gauge can simply be
turned off during the experiment, ion counts from the pump can be
reduced by adding appropriate shielding. The third potential source
of background ions and electrons is the ionisation by stray laser
light of the Rb atoms covering the interior of the vacuum chamber.

Background ions and electrons can be prevented from reaching the
detectors by using suitable ion optics, such as the apparatus
presented in \cite{ionoptics}. In this apparatus four electrodes are
used to accelerate and focus ions and guide them to the detector.
The ion trajectories were calculated using standard software and it
was found that only a volume of \mbox{$\thicksim1$ cm$^3$} is imaged
onto the detector. Ions produced outside of this volume are not
detected and it is estimated that background counts can be almost
entirely suppressed.

From Appendix \ref{app:Nf}, the uncertainty in the number of false
coincidences can also be made very small. From Eqs (\ref{eq:etadet})
and (\ref{eq:sigetai}), neglecting background counts and false
coincidences, and for $p_r\approx1$, the average relative
uncertainty in the ion detection efficiency is
\begin{equation} \label{eq:sigmaoneta} \frac{\sigma_{\eta_{det}}}{\eta_{det}}=
\sqrt{\frac{1-\eta_{det}}{\eta_{det} \eta_{det,e} N_I}}.
\end{equation}  For calibration parameters \mbox{$\dot{N}_I=10^5$ s$^{-1}$}, $\eta_{det}=\eta_{det,e}=0.8$ and a
calibration time of \mbox{$\tau_{cal}=10$ s}, this expression is
accurate to within 1\% for background electron and ion rates of up
to $10^3$ s$^{-1}$. Equation (\ref{eq:sigmaoneta}) shows that this
calibration method works best for large atom numbers and high
overall ion and electron detection efficiencies.

In order to test the values for $\eta_{det}$ and
$\sigma_{\eta_{det}}$ predicted by the calibration equations,
numerical simulations of calibration experiments were carried out.
These simulations included all the relevant features of a real
calibration, such as

\begin{itemize}
  \item ionisation following Poissionian statistics
  \item detection of ions and electrons subject to their respective
detection efficiencies
  \item background electrons and ions
  \item resolution times of the detectors
  \item generation of a coincidence window from the lists of electron
and ion arrival times
  \item use of the coincidence window to determine the
number of coincidences
\end{itemize}

Each simulation generated electron, ion and coincidence counts for
both ``real'' and background electrons and ions.  Then, as would be
done in a real calibration experiment, the corrected counts were
calculated using the estimated numbers of background electrons,
background ions and false coincidences. An inferred value and an
inferred uncertainty of the ion detection system efficiency
$\eta_{det}$ were then found using Eqs (\ref{eq:etadet}) and
(\ref{eq:sigetai}).

\begin{figure}
\includegraphics[width=3.5in]{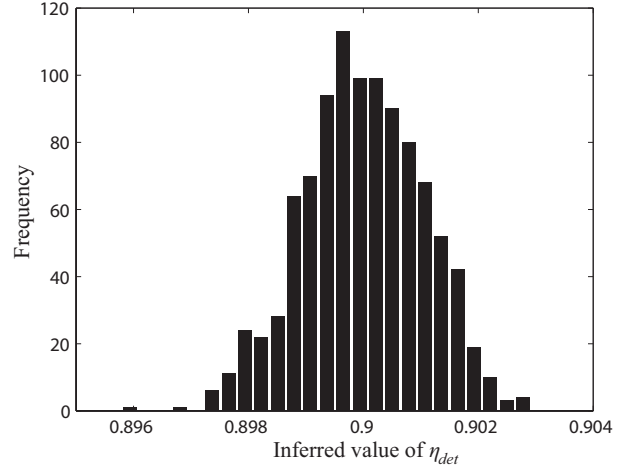} \caption{
Histogram of the inferred values of the ion detection system
efficiency $\eta_{det}$ for 1000 runs of the calibration simulation
with \mbox{$\eta_{det}=0.9$}, electron detection system efficiency
$\eta_{det,e}=0.9$, ionisation rate \mbox{$\dot{N}_I=10^4$
s$^{-1}$}, calibration time \mbox{$\tau_{cal}=10$ s}, pulse
resolution time \mbox{$\tau_r=10$ ns} and, background electron and
ion detection rates \mbox{$\dot{N}_{ib}=\dot{N}_{eb}=10^2$ s$^{-1}$}
respectively.} \label{fig:histogram}
\end{figure}

Excellent agreement was found between the theory and simulation for
a wide range of values. Figure \ref{fig:histogram} shows a histogram
of the inferred ion detection efficiency for 1000 runs of the
simulation with $\eta_{det}=\eta_{det,e}=0.9$,
\mbox{$\dot{N}_I=10^4$ s$^{-1}$}, \mbox{$\tau_{cal}=10$ s},
\mbox{$\tau_r=10$ ns} and $\dot{N}_{ib}=\dot{N}_{eb}$ \mbox{$=10^2$
s$^{-1}$}. The inferred values of $\eta_{det}$ have a mean of
0.899993 in excellent agreement with the actual value of 0.9.  The
standard deviation of the inferred values was \mbox{$1.047 \times
10^-3$}, also in excellent agreement with the average theoretical
value of $\sigma_{\eta_{det}}$, which for these parameters is $1.049
\times 10^{-3}$.

To determine the values of $\eta_i$ and $\sigma_{\eta_i}$ that apply
in a counting experiment, the calibrated values of $\eta_{det}$ and
$\sigma_{\eta_{det}}$ must be corrected for the value of the ion
temporal resolution probability $p_r$ that applies for the ion
detection rate used in the experiment. $\eta_i$ is found from Eq.
(\ref{eq:eta}) and $\sigma_{\eta_i}=\sigma_{\eta_{det}}$ for
$p_r\approx1$.  For the case of ionisation at a constant rate in
atoms per second, $p_r$ is given by Eq. (\ref{eq:pres3}) in Appendix
\ref{app:pr}.  If ionisation occurs at a constant per atom rate
(i.e. ionising laser powers are held constant), we use the effective
ion resolution probability $p'_r$, given by Eq. (\ref{eq:pdashr}).
It should be noted that for the calibration to be valid, the ion
collection efficiency $\eta_{coll}$ must remain constant between
calibration and experiment. This could be achieved by calibrating
using cold atoms in the same trap as used in the experiment.  To
achieve the lowest possible uncertainty in the atom count,
$\eta_{coll}$ should be unity, which can in principle be achieved
with well-designed ion optics.

\section{Trap loss and optimal ionisation rate}

\begin{figure}
\includegraphics[width=3.8in]{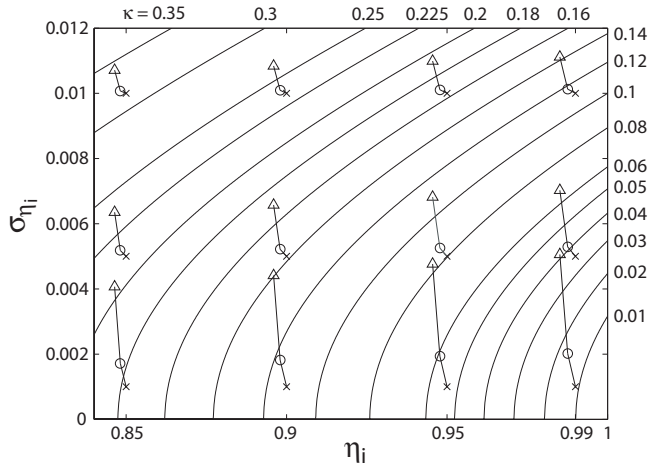} \caption{Contour plot of the normalised count uncertainty $\kappa$ showing how $\kappa$ increases as the
ionisation time $\tau_I$ decreases, for various combinations of
initial values of the overall ion detection efficiency $\eta_i$ and
its uncertainty $\sigma_{\eta_i}$. Crosses, circles and triangles
correspond to $\tau_I \rightarrow \infty$, $\tau_I=2$ ms and
$\tau_I=1$ ms respectively.} \label{fig:tauIplot}
\end{figure}

In an atom counting experiment it is important to know how the rate
of ionisation affects the normalised count uncertainty $\kappa$. The
greater the rate of ionisation, the smaller the value of $p_r$ and
the larger its uncertainty. This in turn decreases $\eta_i$ and
increases $\sigma_{\eta_i}$, resulting in an increased value of
$\kappa$. However in most experiments it would generally be
desirable to ionise a cloud as quickly as possible without overly
affecting the count accuracy. Figure \ref{fig:tauIplot} shows how
$\kappa$ is affected by the ionisation time $\tau_I \equiv
N_I/\dot{N}_{I} $, for $N_I=10^3$ and \mbox{$\tau_r=5$ ns}, assumed
to be known with a \mbox{10\%} uncertainty. The crosses correspond
to an infinitely long ionisation (in which case $\eta_i=\eta_{det}$)
and the circles and triangles correspond to ionisation times of 2
and 1 ms respectively.  It can be seen that down to a certain
ionisation time, $\kappa$ is close to its minimum value.  However as
the ionisation time decreases, $\kappa$ begins to increase rapidly.

If there is negligible trap loss, the number of atoms ionised $N_I$
is the same as the number of atoms originally in the trap $N$, and
the inferred atom number $N_{inf}$ is the same as the inferred ion
number $N_{I,inf}$. If there is trap loss however, we have to
correct the inferred ion number to obtain an inferred atom number.
In the remainder of this section we show how this is achieved and
how the uncertainty in the inferred atom number is affected by trap
loss. We also find the optimal ionisation rate which minimises the
uncertainty in the inferred atom number due to both trap loss and
imperfect ion temporal resolution.

To calculate the atom count corrected for trap loss, we assume a
constant loss rate $R_{l}$ and that ionisation occurs at a
constant rate $R_I$ (both rates are per atom). The respective time
dependent probabilities of an atom being in the trap and of having
been ionised follow
\begin{eqnarray} \label{eq:PdotT} \dot{P}_T & = & -(R_I+R_{l})P_T, \\ \label{eq:PdotI} \dot{P}_I & = & R_I
P_T.
\end{eqnarray} The probability that an atom is ionised rather than
lost is \begin{equation} \label{eq:p_I} p_I= \int_{0}^{\infty}
\dot{P}_I dt = \frac{R_I}{R_I+R_{l}}.
\end{equation}The total number of atoms ionised $N_I$ is described by a binomial
distribution, which again can be approximated by a normal
distribution with mean $p_I N$ and standard deviation
$\sqrt{p_I(1-p_I)N}$ \cite{knoll79}. The original number of atoms in
the trap can then be inferred to be
\begin{equation} \label{eq:Ninf} N_{inf} =
\left(1+\frac{R_{l}}{R_I}\right)N_{I,inf},
\end{equation} with a statistical uncertainty of \begin{equation}
\sigma_{N_{inf},stat} = \sqrt{\frac{R_{l}}{R_I}
\left(1+\frac{R_{l}}{R_I}\right)N_{I,inf}}. \end{equation}
 Due to uncertainties in $R_{l}$, $R_I$ and $N_{I,inf}$, we get extra terms in the
uncertainty in $N_{inf}$ and we find
\begin{multline} \label{eq:sigNinf} \sigma_{N_{inf}}^2=\left[\frac{N_{I,inf}
\sigma_{R_{l}}}{R_I}\right]^2+\left[\frac{R_{l} N_{I,inf}
\sigma_{R_I}}{R_I^2}\right]^2+\\
\left[\left(1+\frac{R_{l}}{R_I}\right)\sigma_{N_{I,inf}}\right]^2+\frac{R_{l}}{R_I}
\left[1+\frac{R_{l}}{R_I}\right]N_{I,inf}, \end{multline} with
$N_{I,inf}$ and $\sigma_{N_{I,inf}}$ given by Eqs
(\ref{eq:NIinfbar}) and (\ref{eq:NIinfsig}). Thus in order to infer
the atom number and its uncertainty, we need to know $R_I$ and $R_l$
and their uncertainties in addition to $N_i$, $\eta_i$, and
$\sigma_{\eta_i}$. We now consider how these values can be measured
in a given experiment.

The sum of $R_I$ and $R_{l}$ can be found by plotting
\mbox{$\ln[N_i-N_i(t)]$} against time, where $N_i(t)$ is the
cumulative ion count. The slope is then $-(R_I+R_{l})$. The value of
$R_{l}$ (and hence of $R_I$) may be found by plotting $R_I+R_{l}$
against the intensity of the 778 nm laser for a number of runs.
$R_{l}$ is given by the extrapolated $y$-intercept of this plot.
(Extrapolation is required because $R_l$ cannot be measured by
counting ions if no ionisation occurs. However, $R_l$ could be
measured directly by absorption imaging at successive times in the
absence of 778 nm light.)

\begin{figure}
\includegraphics[width=3.5in]{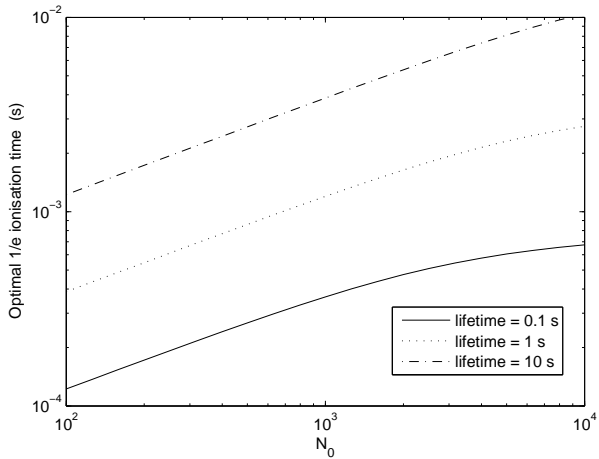} \caption{The optimal 1/e time required to
ionise the cloud is shown as a function of the atom number and the
loss rate from the trap.} \label{fig:ItimevsN}
\end{figure}

\begin{figure}
\includegraphics[width=3.5in]{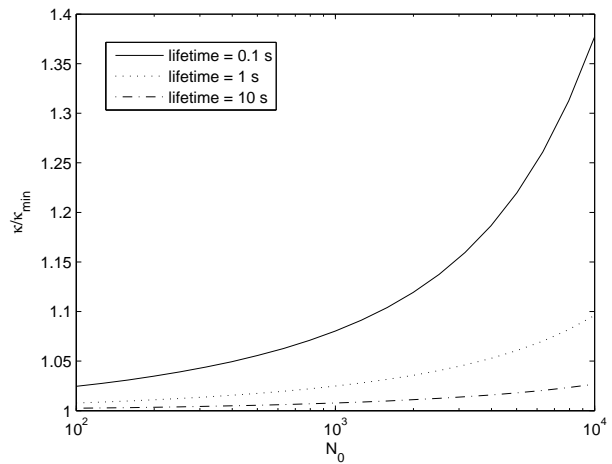} \caption{The minimum achievable value of the relative count uncertainty $\kappa$ expressed relative to
 $\kappa_{min}$ (the value for zero loss) is shown as a function of atom number and trap lifetime.
 For the given values of the overall ion detection efficiency $\eta_i$ and its uncertainty $\sigma_{\eta_i}$, $\kappa_{min}\approx0.11$.}
\label{fig:kvsN}
\end{figure}

The optimal ionisation rate is that which gives the lowest value of
$\kappa$ for given values of $R_l$, $\eta_{det}$, $\tau_r$ and $N$.
In Fig. \ref{fig:ItimevsN}, the optimal value of $1/R_I$ is plotted
as a function of $N$ for various values of $1/R_l$, for
$\eta_{det}=0.9$, $\sigma_{\eta_i}=0$, $\tau_r=10$ ns and assuming a
10\% uncertainty in the measured values of $R_l$ and $R_I$. The
corresponding minimum values of $\kappa/\kappa_{min}$ are plotted in
Fig. \ref{fig:kvsN}, where $\kappa_{min}$ is the value of $\kappa$
for zero trap loss. It can be seen that reasonably sized atom clouds
($10^3-10^4$ atoms) can be counted in times on the order of 10 ms
with close to the theoretical minimum count uncertainties for
realistic trap lifetimes.

An improvement on the uncertainties resulting from ionising with
constant per atom probability would be obtained by increasing the
778 nm laser power during the ionisation in such a way as to obtain
a contant ionisation rate in atoms/s.  This would mean that the
cloud would be ionised faster with less trap loss.  The time
dependent rate equations describing this scheme could be integrated
numerically to determine the probability that an atom is ionised
rather than lost, from which the original atom number and its
uncertainty could be inferred.

\section{Discussion}\label{sect:discussion}

We now consider the application of our atom counting scheme to the
characterisation of unknown atomic samples. To determine the number
of atoms in a single cloud, one would simply count a certain number
of ions $N_i$ and use the known calibration data to obtain an
inferred atom number and uncertainty $N_{inf} \pm \sigma_{N_{inf}}$
from Eqs (\ref{eq:NIinfbar}) and (\ref{eq:NIinfsig}).  In the case
that trap losses are not negligible, Eqs (\ref{eq:Ninf}) and
(\ref{eq:sigNinf}) are also required. If however we wish to know the
atom number variance in a specific situation we need to analyse a
number of repeated measurements. As an example, consider a sample of
$n$ measurements on atom clouds generated in an identical manner
which, under ideal (noise free) circumstances, would yield identical
numbers of atoms (for example using a quantum tweezer to extract
atoms from a BEC \cite{diener02,chuu05}). Each experiment yields a
measure $N_{inf,j} \pm \sigma_{N_{inf,j}}$ where $j = 1, 2 .. n$.
The quantities of interest are the mean atom number, the atom number
variance and the uncertainties in these values.

The population mean can be estimated from the mean of all inferred
atom numbers $\bar{N}$ and has an uncertainty of
$\sqrt{s_T^2/n+(\sigma_{\eta_i}\bar{N}/\eta_i)^2}$ where $s_T^2$ is
the bias-corrected sample total variance and is given by
\mbox{$s^2_T=1/(n-1)\sum(N_{inf,j}-\bar{N})^2$}. The true atom
number variance $\sigma_A^2$ is estimated by the difference of the
total variance of the measurements $s_T^2$ and the variance
associated with the binomial statistics of the ion counting
$\sigma_m^2 \cong \bar{N} (1-\eta_i)/\eta_i$. Note that the
expression for $\sigma_m^2$ does not include a term due to
$\sigma_{\eta_i}$ as in Eq. (\ref{eq:NIinfsig}), since we assume
that any error in the detection efficiency (calibration) will be the
same for each measurement. This leads to a systematic error in the
atom count and does not contribute to the variance $\sigma_m^2$.

The degree of number squeezing of the atomic population is described
by the Fano factor $F_A  =\sigma^2_A / \bar{N}$ \cite{fano47}. A
Poissonian sample has $F_A = 1$ and a sub-Poissonian
(number-squeezed) sample has $F_A < 1$. We can define a total Fano
factor $F_T = s_T^2 / \bar{N}$ and measurement Fano factor $F_m =
\sigma_m^2 / \bar{N}$, which accounts for the fluctuations due to
the binomial counting of ions. The population Fano factor is then
given by $F_A = F_T - F_m$, and it can be shown that for
$\sigma_{\eta_i} < 5$\% and $\bar{N} > 5$ the uncertainty in $F_A$
is well approximated by $\sigma_{F_A} = \sqrt{2/n}(F_A + F_m)$.
Knowing $F_m$ from the calibration it is possible to determine how
many measurements are required to obtain a suitably small
uncertainty $\sigma_{F_A}$ in the measured (normalised) atomic
variance $F_A$.

\section{Conclusion}

We have analysed photoionisation and ion detection as a means of
achieving accurate counting of ultra-cold atoms.  The scheme relies
on a three-photon, cw ionisation process and can also be used for
efficient single atom detection with high spatio-temporal
resolution.  To count large numbers of atoms, one of the lasers acts
as an optical dipole trap for the target atoms. We have shown that a
well calibrated ion detector with realistic efficiencies can achieve
single shot atom counts with sub-Poissonian accuracies for atomic
samples containing many thousands of atoms. This is potentially
useful for studies of the quantum properties of cold atomic gases.
The variance of an ensemble of systems can also be characterised in
a manner that only weakly depends on uncertainties in the absolute
detection efficiency. Similarly, this scheme could allow direct
probing of atom number correlations in molecular down conversion and
four wave mixing experiments which are predicted to show number
difference squeezing.

This work was supported by the Australian Research Council,
Landesstiftung Baden-W\"{u}rttemberg and the European Union
(MRTN-CT-2003-505032).

\appendix

\section{Table of Symbols} \label{app:symbols}

\begin{longtable}{p{.15\linewidth} p{.75\linewidth}}
\hline\\$N$ & number of atoms originally in the trap\\
$N_{inf}$ & inferred no. of atoms originally in the trap\\
$\sigma_{N_{inf}}$ & uncertainty in the inferred atom number \\
$\sigma_{N_{inf,stat}}$ & statistical uncertainty in the inferred atom no.\\
$N_I$ & number of atoms ionised\\
$N_{I,inf}$ & inferred number of atoms ionised\\
$\sigma_{N_{I,inf}}$ & uncert. in the inferred no. of atoms
ionised\\
$\dot{N}_I$ & ionisation rate (s$^{-1})$\\
$\eta_i$ & overall ion detection efficiency\\
$\sigma_{\eta_i}$ & uncertainty in the overall ion detection eff.\\
$\eta_e$ & overall electron detection system efficiency\\
$\eta_{det}$ & ion detection system efficiency\\
$\sigma_{\eta_{det}}$ & uncert. in the ion detection system eff.\\
$\eta_{det,e}$ & electron detection system efficiency\\
$\eta'_{det}$ & ion detector efficiency\\
$\eta_{coll}$ & ion collection efficiency\\
$p_r$ & ion pulse temporal resolution probability\\
$p'_r$ & effective ion pulse temporal resolution prob.\\
$p_{r,e}$ & electron pulse temporal resolution probability\\
$\kappa$ & normalised count uncertainty\\
$\kappa_{min}$ & normalised count uncertainty for zero trap loss\\
$\tau_r$ & pulse resolution time (s)\\
$\tau_{cal}$ & calibration time (s)\\
$\tau_I$ & ionisation time (s)\\
$\tau_{win}$ & coincidence window length (s)\\
$\tau$ & interval between successive ion detections (s)\\
$N_i$ & number of ions counted\\
$N_i(t)$ & cumulative ion count\\
$\dot{N}_i$ & ion detection rate (s$^{-1}$)\\
$\dot{N}^*_i$ & ion detection rate as $\tau_r\rightarrow0$
(s$^{-1}$)\\
$\dot{N}_{i,cal}$ & ion detection rate in a calibration (s$^{-1}$)\\
$N_e$ & number of electrons counted\\
$N_c$ & number of coincidences counted\\
$N'_i$ & corrected ion count\\
$N'_e$ & corrected electron count\\
$N'_c$ & corrected coincidence count\\
$\bar{N}_{ib}$ & mean no. of background ions counted\\
$\bar{N}_{eb}$ & mean no. of background electrons counted\\
$\sigma_{eb}$ & uncertainty in the background electron count\\
$\bar{N}_f$ & mean net no. of false coincidences counted\\
$\sigma_{f}$ & uncertainty in the false coincidence count\\
$p_f$ & probability of a given unpartnered ion being in the window
of an unpartnered electron\\
$p_w$ & probability of a true coincidence being outside the coincidence window\\
$\sigma_{int}$ & standard deviation of the time intervals between
electrons and their corresponding ions\\
$R_l$ & trap loss rate per atom (s$^{-1}$)\\
$\sigma_{R_l}$ & uncertainty in $R_l$ (s$^{-1})$\\
$R_I$ & ionisation rate per atom (s$^{-1})$\\
$\sigma_{R_I}$ & uncertainty in $R_I$ (s$^{-1})$\\
$P_T$ & probability of an atom being in the trap as a function of
time\\
$P_I$ & probability of an atom being ionised as a function of time\\
$p_I$ & probability of an atom being ionised as
$t\rightarrow\infty$\\
$f_{int}(\tau)$ & probability density function of the time interval
between successive detections\\
$f_a(t)$ & probability density function of arrival times of ions at
the detector\\
$\beta$ & $\tau_r \eta_{det} N R_I$\\
$n$ & number of atom number measurements\\
$\bar{N}$ & mean of inferred atom numbers\\
$s^2_T$ & bias-corrected sample atom number total variance\\
$N_{inf,j}$ & inferred atom number in the $j^{th}$ measurement\\
$\sigma_{N_{inf,j}}$ & uncertainty in the inferred atom number in the $j^{th}$ measurement\\
$\sigma^2_A$ & atom number population variance\\
$\sigma^2_m$ & variance in the atom counts due to the binomial
statistics of the measurement\\
$F_A$ & Fano factor of a population of atomic clouds\\
$\sigma_{F_A}$ & uncertainty in the Fano factor of a population of
atomic clouds\\
$F_T$ & Fano factor of a sample of inferred atom numbers\\
$F_m$ & measurement Fano factor\\
\\
 \hline\end{longtable}

\section{Net number of false coincidences} \label{app:Nf}

In this appendix we derive the required correction to a coincidence
count to account for false coincidences and uncounted true
coincidences. In a given calibration there will be $N_i-N_c$
unpartnered ions and $N_e-N_c$ unpartnered electrons. (We assume
negligible background ion and electron numbers.) The probability of
a given unpartnered ion being in the window of an unpartnered
electron is well approximated by
\begin{equation} p_f=\left(N_e-N_c\right)\frac{\tau_{win}}{\tau_{cal}},
\end{equation} where $\tau_{win}$ is the window length and $\tau_{cal}$ is
the duration of the calibration.

The probability of a true coincidence being outside the window is
\begin{equation} p_w =
1-\mathrm{erf}\left(\frac{\tau_{win}}{2\sqrt{2}\sigma_{int}}\right),
\end{equation} where $\sigma_{int}$ is the the standard deviation of the time intervals between electrons and
their corresponding ions.

The net number of false coincidences counted in a given calibration
is then
\begin{equation} \label{eq:Nf} N_f = \bar{N}_f \pm \sigma_f, \end{equation} where
\begin{equation} \bar{N}_f = p_f\left(N_i-N_c\right) -
p_w N_c,
\end{equation} and \begin{equation} \sigma_f = \sqrt{p_f\left(1-p_f\right)\left(N_i-N_c\right)+p_w\left(1-p_w\right)N_c}\end{equation}

\section{Calculation of $p_r$ and $p'_r$} \label{app:pr}

As the overall ion detection efficiency $\eta_i$ is ionisation rate
dependent, relating the ion detection efficiency measured in a
calibration to the ion detection efficiency which applies in an
experiment requires finding an expression for $p_r$ in terms of the
ionisation rate [see Eq. (\ref{eq:eta})]. We consider two cases:
constant ionisation rate in atoms per second, and constant per atom
ionisation rate.  We look first at the former.

We assume the ion detection system has some resolution time
$\tau_{r}$, such that if an ion arrives at the detector within
$\tau_{r}$ of the previous ion it will not be counted. We define
$\dot{N^*_i}$ as the average rate (ions/s) at which ions are
detected as $\tau_r\rightarrow 0$.  Assuming a constant average
detection rate, the detections follow Poissonian statistics, and the
time interval between successive detections $\tau$, is described by
a negative exponential probability density function \cite{knoll79}:
\begin{equation} \label{eq:intpdf} f_{int}(\tau)= \dot{N}^*_i
\mathrm{e}^{-\dot{N}^*_i \tau} \qquad \textrm{for } \tau \geq 0.
\end{equation}

The probability that an ion arriving at the detector is resolvable
from the previous ion is
\begin{equation} \label{eq:pres}
p_{r}=1-\int_0^{\tau_r} f_{int}(\tau) d\tau.
\end{equation} From Eqs (\ref{eq:intpdf}) and (\ref{eq:pres}) we find
\begin{equation} \label{eq:pres2} p_{r} = \mathrm{exp}(-\tau_r
\dot{N}^*_i). \end{equation} The rate at which ions are actually
detected is
\begin{equation} \label{eq:Ndoti} \dot{N}_i = p_r \dot{N}^*_i.
\end{equation} The number of ions counted during a calibration
period $\tau_{cal}$ is described by a Poisson distribution:
$N_i=\dot{N}_i\tau_{cal}\pm \sqrt{\dot{N}_i \tau_{cal}}$. If $N_i$
ions are counted in time $\tau_{cal}$, the ion detection rate in the
calibration is inferred to be
\begin{equation} \label{eq:Ndoti2} \dot{N}_{i,cal}=\frac{N_i}{\tau_{cal}}
\pm \frac{\sqrt{N_i}}{\tau_{cal}}.
\end{equation}From Eqs (\ref{eq:pres2}) and (\ref{eq:Ndoti})
we find \begin{equation} \label{eq:Ndoti'} \dot{N}^*_i = (1+\tau_r
\dot{N}_i) \dot{N}_i + O\left((\tau_r \dot{N}_i)^3\right).
\end{equation} Substituting Eq. (\ref{eq:Ndoti'}) into Eq. (\ref{eq:pres2})
and neglecting the third order term (which results in a relative
error in $p_r$ of less than $10^{-9}$ for $\tau_r \dot{N}_i <
10^{-3}$) we get
\begin{equation} \label{eq:pres3} p_r=1-\tau_r \dot{N}_i -
\frac{1}{2}(\tau_r \dot{N}_i)^2.
\end{equation}

Now we consider the case of ionisation with a constant per atom
ionisation rate.  In this case we need to use an effective ion
resolution probability  $p'_r$ because by ionising with a constant
per atom rate, the ionisation rate in ions per second is a function
of time and hence so is $p_r$. The effective ion resolution
probability is given by
\begin{equation} p'_r=\int_{0}^{\infty} p_r(t) f_a(t) dt,
\end{equation} where from Eqs (\ref{eq:pres2}), (\ref{eq:PdotT}) and
(\ref{eq:PdotI})
\begin{equation} p_r(t)= \exp\left(-\beta \mathrm{e}^{-(R_I+R_l)t}\right), \end{equation}
 where $\beta=\tau_r \eta_{det} N R_I$ and
 \begin{equation} f_a(t)=(R_I+R_l)\mathrm{e}^{-(R_I+R_l)t} \end{equation}
is the probability density function of arrival times of ions at the
detector.  It can be shown that, neglecting terms of $O(\beta^3)$,
\begin{equation} \label{eq:pdashr}
p'_r=1-\frac{1}{2}\beta+\frac{1}{3}\beta^2.
\end{equation}


\begin{thebibliography}{99}
\bibitem{adams} C. S. Adams, M. Siegel and J. Mlynek, Phys. Rep.
\textbf{240}, 143 (1994).
\bibitem{Anderson95} M. Anderson, J. R. Ensher, M. R. Matthews,
C. E. Wieman and E. A. Cornell, Science {\bf 269}, 198 (1995).
\bibitem{Bradley95}
C. C. Bradley, C. A. Sackett, J. J. Tollett and  R. G. Hulet, Phys.
Rev. Lett. {\bf 75}, 1687 (1995).
\bibitem{Davis95}
K. B. Davis, M-O.Mewes, M. R. Andrews, N. J. van Druten, D.S.
Durfee, D.M. Kurn and W. Ketterle, Phys. Rev. Lett. {\bf 75}, 3969 (1995).

\bibitem{andrews97} M. R. Andrews, C. G. Townsend, H. J. Miesner,
D. S. Durfee, D. M. Kurn, and W. Ketterle, Science {\bf 275}, 637
(1997).
\bibitem{mewes97} M. -O. Mewes, M. R. Andrews, D. M. Kurn, D. S. Durfee,
C. G. Townsend, and W. Ketterle, Phys. Rev. Lett. {\bf 78}, 582
(1997).
\bibitem{bloch99} I. Bloch, T. W. H\"{a}nsch and T. Esslinger,
Phys. Rev. Lett. {\bf 82}, 3008 (1999).
\bibitem{hagley99}
E. W. Hagley, L. Deng, M. Kozuma, M. Trippenbach, Y. B. Band, M.
Edwards, M. Doery, P. S. Julienne, K. Helmerson, S. L. Rolston, and
W. D. Phillips, Phys. Rev. Lett. {\bf 83}, 3112 (1999).
\bibitem{lecoq01}Y. Le Coq, J. H. Thywissen, S. A. Rangwala, F. Gerbier,
S. Richard, G. Delannoy, P. Bouyer and A. Aspect, Phys. Rev. Lett.
{\bf 87}, 170403 (2001).
\bibitem{moelmer03} K. M{\o}lmer, New J. Phys.
{\bf 5}, 55 (2003).
\bibitem{burt97} E. A. Burt, R. W. Ghrist, C. J. Myatt, M. J. Holland,
E. A. Cornell, and C. E. Wieman, Phys. Rev. Lett. {\bf 79}, 337
(1997).
\bibitem{tolra04}
B. Laburthe Tolra, K. M. O'Hara, J. H. Huckans,
W. D. Phillips, S. L. Rolston, and J. V. Porto,
Phys. Rev. Lett. \textbf{92}, 190401 (2004).
\bibitem{kinoshita05}
T. Kinoshita, T. R. Wenger and D. S. Weiss,
Phys. Rev. Lett. \textbf{95}, 190406 (2005).
\bibitem{esteve06}
J. Esteve, J-B. Trebbia, T. Schumm, A. Aspect, C. I. Westbrook and
I. Bouchoule, cond-mat/0510397.
\bibitem{schellekens05}
M. Schellekens, R. Hoppeler, A. Perrin, J. Viana Gomes, D. Boiron, A. Aspect and
C. I. Westbrook, Science \textbf{310} 648 (2005).
\bibitem{ottl05} A. \"{O}ttl, S. Ritter, M. K\"{o}hl and T.
Esslinger, Phys. Rev. Lett. {\bf 95}, 090404 (2005).
\bibitem{javanainen99} J. Javanainen and M. Y. Ivanov, Phys. Rev. A
{\bf 60}, 002351 (1999).
\bibitem{orzel}
C. Orzel, A. K. Tuchman, M. L. Fenselau, M. Yasuda and M. A.
Kasevich, Science \textbf{291}, 238 (2001).
\bibitem{greiner02}
M. Greiner,  O. Mandel, T. Esslinger, T. W. H\"{a}nsch and I. Bloch,
Nature \textbf{415}, 39 (2002).
\bibitem{hadzibabic04}
Z. Hadzibabic, S. Stock, B. Battelier, V. Bretin and J. Dalibard,
Phys. Rev. Lett. {\bf 93},  180403 (2004).
\bibitem{Pu00}
H. Pu and P. Meystre, Phys. Rev. Lett. \textbf{85} 3987, (2000).
\bibitem{Duan00}
L.-M. Duan, A. S\o rensen, J.~I. Cirac, and P. Zoller,
Phys. Rev. Lett. \textbf{85} 3991, (2000).
\bibitem{Sorensen01}
A. S\o rensen, L.-M Duan, J. I. Cirac, P. Zoller,
Nature \textbf{409}, 63 (2001).
\bibitem{roberts02}
D. C. Roberts, T. Gasenzer and K. Burnett,
J. Phys. B \textbf{35}, L113 (2002).
\bibitem{kherunts02}
K. V. Kheruntsyan and P. D. Drummond, Phys. Rev. A \textbf{66},
 031602(R) (2002).
\bibitem{kherunts05}
K. V. Kheruntsyan, Phys. Rev. A \textbf{71}, 053609 (2005).
\bibitem{Deng1999} L. Deng, E. W. Hagley, J. Wen, M. Trippenbach, Y. Band, P. S. Julienne,
 J. E. Simsarian, K. Helmerson, S. L. Rolston and W. D. Phillips,
Nature \textbf{398}, 218 (1999).
\bibitem{campbell06}
G.~K. Campbell, J. Mun, M. Boyd, E.~W. Streed, W. Ketterle and D.~E Pritchard,
Phys. Rev. Lett. \textbf{96}, 020406 (2006).
\bibitem{inouye99} S. Inouye, T. Pfau, S. Gupta, A.P. Chikkatur, A. Görlitz, D.E.
Pritchard, and W. Ketterle, Nature {\bf 402}, 641-644 (1999).
\bibitem{greiner05} M. Greiner, C. A. Regal, J. T. Stewart, and D. S. Jin
Phys. Rev. Lett. {\bf 94}, 110401 (2005).
\bibitem{folling05} S. F\"{o}lling, F. Gerbier, A. Widera, O. Mandel,
T. Gericke and I. Bloch, Nature {\bf 434}, 481 (2005).
\bibitem{diener02} R. B. Diener, B. Wu, M. G. Raizen and Q. Niu, Phys.
Rev. Lett. \textbf{89}, 070401 (2002).
\bibitem{chuu05} C. S. Chuu, F. Schreck, T. P. Meyrath, J. L. Hanssen,
G. N. Price and M. G. Raizen, Phys. Rev. Lett. \textbf{95}, 260403
(2005)
\bibitem{frese00} D. Frese, B. Ueberholz, S. Kuhr, W. Alt, D. Schrader,
V. Gomer, and D. Meschede, Phys. Rev. Lett. {\bf 85}, 3777 (2000).
\bibitem{schlosser01} N. Schlosser, G. Reymond, I. E. Protsenko and P. Grangier,
Nature {\bf 411}, 1024 (2001).
\bibitem{teper06} I. Teper, Y.-J. Lin, and V. Vuleti\'{c},
cond-mat/0603675 (2006).
\bibitem{pinske00} P.W.H. Pinkse, T. Fischer, P. Maunz, and G.
Rempe,
Nature {\bf 404}, 365-368 (2000).
\bibitem{hood00} C. J. Hood, T. W. Lynn, A. C. Doherty, A. S. Parkins,
and H. J. Kimble, Science \textbf{287}, 1457 (2000).
\bibitem{mabuchi02} H. Mabuchi and A. C. Doherty,
Science {\bf 298}, 1372, (2002).
\bibitem{horak03} P. Horak, B. G. Klappauf, A. Haase, R. Folman, J. Schmiedmayer,
P. Domokos, E. A. Hinds, Phys. Rev. A {\bf 67}, 043806 (2003).
\bibitem{lye03} J. E. Lye, J. J. Hope and J. D. Close,
Phys. Rev. A {\bf 67}, 043609 (2003).
\bibitem{baldwin05} K. G. H. Baldwin,
Contemp. Phys. {\bf 46}, 105 (2005).
\bibitem{ruschhaupt04} A Ruschhaupt, B Navarro and J G Muga
J. Phys. B: At. Mol. Opt. Phys. {\bf 37}, L313 (2004).
\bibitem{ristrophe05} T. Ristroph, A. Goodsell, J.A.
Golovchenko, and L.V. Hau,  Phys. Rev. Lett. \textbf{94}, 066102
(2005).
\bibitem{frank96} M. Frank, C. A. Mears, S. E. Labov, W. H. Benner,
D. Horn, J. M. Jaklevic and A. T. Barfknecht, Rapid Communications
in Mass Spectrometry, \textbf{10}, 1946 (1996).
\bibitem{duncan01} B. C. Duncan, V. Sanchez-Villicana, P. L.
Gould and H. R. Sadeghpour, Phys. Rev. A, \textbf{63}, 043411
(2001).
\bibitem{ciampini02} D. Ciampini, M. Anderlini, J. H. M\"{u}ller, F.
Fuso, O. Morsch, J. W. Thomsen and E. Arimondo, Phys. Rev. A
\textbf{66}, 043409 (2002).
\bibitem{ionoptics} S. Kraft, PhD thesis, Universit\"{a}t
T\"{u}bingen (2006). Details will be published elsewhere.
\bibitem{knoll79}G. Knoll, Radiation Detection and
Measurement, (1979).
\bibitem{ncrp}NCRP Report 58, A Handbook of
Radioactivity Measurements Procedures, second ed., 76 (1985).
\bibitem{ortec}Ortec 100-ps Time Digitizer / MCS Model 9395.
\bibitem{fano47} U. Fano, Phys. Rev. \textbf{72}, 26 (1947).
\end{thebibliography}
\end{document}